\documentstyle[11pt,amsfonts]{article}

\everymath={\displaystyle}

\newcommand{\be}{\begin{equation}}
\newcommand{\ee}{\end{equation}}

\date{November 2001}
\begin{document}
\begin{center}
{\large {\bf Darboux transformations and hidden quadratic supersymmetry of 
the one-dimensional stationary Dirac equation}}\\ 
\vspace{2cm}
 N. DEBERGH\footnote{email: Nathalie.Debergh@ulg.ac.be}(a), 
A.A. PECHERITSIN\footnote{email: pecher@ido.tsu.ru}(b),\\ 
B.F. SAMSONOV\footnote{email: samsonov@phys.tsu.ru}(b) and 
B. VAN DEN BOSSCHE\footnote{email: bvandenbossche@ulg.ac.be}(a)\\ \vspace{1cm}
(a){\it Fundamental Theoretical Physics,\\
Institute of Physics (B5),\\
University of Li\`ege,\\
B-4000 LIEGE  (Belgium)}\\ \vspace{0.5cm}

(b){\it Department of Quantum Field Theory,\\
Tomsk State University,\\
36 Lenin Ave., \\
634050 TOMSK (Russia)} \\
\end{center}
\vspace{0.5cm}
\begin{abstract}
A matricial Darboux operator intertwining two one-dimensional stationary Dirac 
Hamiltonians is constructed. This operator is such that the potential of 
the second Dirac Hamiltonian as well as the corresponding eigenfunctions 
are determined through the knowledge of only two eigenfunctions of the 
first Dirac Hamiltonian. Moreover this operator together with its adjoint 
and the two Hamiltonians generate a quadratic deformation of the superalgebra 
subtending the usual supersymmetric quantum mechanics. Our developments 
are illustrated on the free particle case and the generalized Coulomb 
interaction. In the latter case, a relativistic counterpart of 
shape-invariance is observed.
\end{abstract}
\newpage

\section{Introduction}
In quantum mechanics, the Schr\" odinger equations which can be solved 
by analytic methods exclusively are rather exceptional. Therefore the 
methods being able to enlarge the number of such equations have attracted 
much attention in recent as well as less recent literature. Three of them 
still remain very popular: the Darboux transformations \cite{1} elaborated in 
1882 within the mathematical framework of Sturm-Liouville differential 
equations, the factorization method introduced by Schr\" odinger \cite{2} in 
1940 and more recently the so-called supersymmetric quantum mechanics 
\cite{3}. 
All of them are more or less based on the same following ideas. 

Let us 
consider the following Schr\" odinger Hamiltonian
\be
H_0 \equiv -\frac{d^2}{dx^2}+V_0(x), \; x \in {\Bbb R} \; {\rm or} \; x \in
{\Bbb R}_0^+
\ee
which can be factorized as follows 
\be
H_0=L^{\dagger}L + \alpha, \; \alpha={\rm constant}
\ee
with
\be
L=\frac{d}{dx} + W(x).
\ee
Then the eigenfunctions of the isospectral 
(up to the eventual creation or loss of one energy) Hamiltonian $H_1$ 
defined by
\be
H_1 \equiv L L^{\dagger} + \alpha = -\frac{d^2}{dx^2}+V_1(x)
\ee
are obtained through the application of $L$ to the eigenfunctions of $H_0$ 
as it is clear from the so-called intertwining relation
\be
L H_0 = H_1 L.
\label{eq5}
\ee
Thus a new Schr\" odinger Hamiltonian $H_1$ has been constructed and it
 is exactly solvable if $H_0$ is.

We remark that the Darboux transformation has two particular features
compared to~\cite{2,3}: 
First, the potential  $W(x)$, written as
\be
W(x)=-\frac{d\ {\rm ln} \psi_0(x)}{dx},
\ee
can be constructed  from a bounded {\it or unbounded} 
non-vanishing eigenfunction $\psi_0(x)$  
of $H_0$ (with eigenvalue $\alpha$). 
Second, the Darboux operator 
$L$ can be extended to  higher order \cite{1,4}.

Here we shall ask for the same kind of developments in the relativistic 
context that is to say search for the operator $L$ intertwining two 
one-dimensional Dirac Hamiltonians. A partial answer has already been given 
in \cite{5} and \cite{6} through the supersymmetrical features 
of specific Dirac 
Hamiltonians. Another one can also be found in \cite{7,8} where a relativistic 
Darboux transformation has been considered but for pseudoscalar potentials 
only. In the following we will not limit ourselves to such a context and 
will give, in Section~\ref{section2}, the extended intertwining operator $L$ 
corresponding to a general self-adjoint potential. This operator is 
constructed from two (known) solutions of the initial Dirac equation 
and gives rise to new exactly solvable Dirac equations. Moreover, in 
Section~\ref{section3}, we will convince ourselves 
from this operator $L$ and its 
adjoint that the underlying superstructure in the relativistic context is a 
quadratic {\it deformation} of the $sqm(2)$ superalgebra, the latter 
being, as well known \cite{3}, the subtending superalgebra of the 
(non-relativistic) supersymmetric quantum mechanics. Finally, in 
Section~\ref{section4}, we will illustrate our statements on two examples: 
The free particle case and the generalized 
Coulomb interaction. For the latter, we observe 
the relativistic counterpart of the so-called shape-invariance \cite{9}, i.e., 
only the values of the parameters introduced in the expression of the 
potentials change.

\section{Intertwining operator for the Dirac equation}
\label{section2}
Let us start with the following one-dimensional Dirac Hamiltonian
\be
h_0 \equiv i \sigma_2 \frac{d}{dx} + v_0(x), \; x \in {\Bbb R} \; {\rm or} 
\; x \in {\Bbb R}_0^+
\label{eq7}
\ee
where $\sigma_2$ is the usual two-by-two Pauli matrix and $v_0$ is real and 
symmetric, i.e.,
\be
v_0(x)=\left( \begin{array}{ll}
  v_{11}^0(x) & v_{12}^0(x)\\
v_{12}^0(x) & v_{22}^0(x)\\
\end{array}\right).
\label{eq8}
\ee
We assume here that $h_0$ is a known exactly solvable Hamiltonian; in other 
words, all its eigenfunctions, the two-component spinors $\psi(x)$, as well as 
the corresponding energies are analytically determined. Let us now search 
for a matricial operator $L$ satisfying the intertwining relation similar
 to~(\ref{eq5}), i.e.,
\be
L h_0 = h_1 L
\label{eq9}
\ee
with
\be
h_1 \equiv i \sigma_2 \frac{d}{dx} + v_1(x),
\label{eq10}
\ee
$v_1(x)$ being at this level the unknown real and symmetric potential. 
The simplest operator $L$ we can consider is
\be
L \equiv A \frac{d}{dx} + B
\label{eq11}
\ee
where $A$ and $B$ are two-by-two matrices with $x$-dependent entries. 
The relations~(\ref{eq9}) and (\ref{eq11}) give the following system
\be
[A,\sigma_2]=0,
\label{eq12}
\ee
\be
[B,\sigma_2] -iA v_0 +iv_1 A - \sigma_2 A_x=0,
\label{eq13}
\ee
\be
A v_{0x} + Bv_0 - v_1B - i\sigma_2 B_x=0,
\label{eq14}
\ee
the notation $A_x$ meaning here 
$$\frac{dA}{dx} \equiv \left( \begin{array}{cc}
\frac{d A_{11}}{dx} & \frac{d A_{12}}{dx}\\
\\
\frac{d A_{21}}{dx} & \frac{d A_{22}}{dx}\\
\end{array}\right).$$
The condition (\ref{eq12}) is equivalent to ask for $A_{11}=A_{22}$ 
and $A_{12}=-A_{21}$. 
The constraint (\ref{eq13}) enables us to fix the potential 
difference $\Delta v \equiv v_1-v_0$
\be
\Delta v= (Av_0-v_0A+i[B,\sigma_2]-i\sigma_2A_x)A^{-1}
\label{eq15}
\ee
up to the assumption of the existence of $A^{-1}$. Finally from (\ref{eq14})
 we can obtain the matrix $B$ or in a simpler way $\sigma$ defined through 
$B \equiv A \sigma$. Indeed, Eq.~(\ref{eq14}) then reads
\be
(v_0-i\sigma_2 \sigma)_x + [\sigma , v_0] +i [\sigma_2 , \sigma] \sigma =0.
\label{eq16}
\ee
We recognize a matrix analogue of the Riccati equation. 
It can be linearized through the substitution
\be
\sigma = - u_x u^{-1}
\label{eq17}
\ee
in order to become
\be
[u^{-1}(v_0 u + i\sigma_2 u_x)]_x=0
\label{eq18}
\ee
which after integration leads to 
\be
h_0 u = i \sigma_2 u_x + v_0 u= u \lambda,
\label{eq19}
\ee
the matrix $\lambda$ being the constant of integration.

This equation~(\ref{eq19}) is thus formally speaking an ordinary Dirac 
one up to the fact that the solution $u$ is not a spinor anymore 
but a two-by-two matrix while the usual energy $E$ has also been replaced 
by a two-by-two matrix $\lambda$.

The next step is to find a convenient $u$ that is a solution of~(\ref{eq19})  
being real (and invertible) in order to ensure the self-adjointness 
of $v_1$ through~(\ref{eq15}). It is ensured in a straightforward manner if
\be
u=(u_1,u_2)\; , \; \lambda= {\rm diag} (\varepsilon_1,\varepsilon_2)
\label{eq20}
\ee
with the spinors $u_1$ and $u_2$ being eigenfunctions 
(not necessarily bounded) of the Dirac Hamiltonian $h_0$
\be
h_0 u_j= \varepsilon_j u_j, \; j=1,2.
\label{eq21}
\ee
Having found $u$, the operator $L$ given in Eq.~(\ref{eq11}) or
\be
L=A\left(\frac{d}{dx}-u_xu^{-1}\right)
\label{eq22}
\ee
as well as the new potential $v_1$ (see Eq.~(\ref{eq15}))
\be
v_1=A\left(v_0+i[\sigma,\sigma_2]-i\sigma_2A^{-1}A_x\right)A^{-1}
\label{eq23}
\ee
are now fixed up to the determination of $A$. This matrix keeps 
arbitrariness: all one knows is that it has to commute with $\sigma_2$. 
For simplicity and comparison with the non-relativistic context, 
we put $A$ equal to the identity matrix. Eqs.~(\ref{eq22}) and~(\ref{eq23}) 
are then simplified as follows
\be
L= \frac{d}{dx}-u_xu^{-1},
\label{eq24}
\ee
\be
v_1=v_0+i[\sigma, \sigma_2].
\label{eq25}
\ee
These results are the relativistic analogues of the usual Darboux 
transformation. We now give another expression in what concerns 
$v_1$, particularly useful for applications. Indeed from~(\ref{eq19}) we have
\be
\sigma=-u_xu^{-1}=i\sigma_2 u \lambda u^{-1} -i\sigma_2 v_0
\label{eq26}
\ee
and therefore
\be
v_1=\sigma_2 v_0 \sigma_2 + u \lambda u^{-1} - 
\sigma_2 u \lambda u^{-1} \sigma_2,
\label{eq27}
\ee
i.e.,
\be
v_1=\sigma_2 v_0 \sigma_2 + 
\frac{\varepsilon_1-\varepsilon_2}{{\rm det} u} \left( \begin{array}{ll}
d_1 & d_2\\
d_2 & -d_1\\
\end{array}\right),
\label{eq28}
\ee
where $d_1 \equiv u_{11}u_{22}+u_{12}u_{21}$, 
$d_2=u_{21}u_{22}-u_{11}u_{12}$, with $u_{ij}$ corresponding to the element
 of the matrix $u$ being at the crossing of the $i^{th}$ lign and
 the $j^{th}$ column.

Let us close this Section by noticing that, by definition, the 
operator $L$ has a non-trivial kernel since $kerL=u$. This implies
 that the action of $L$ to an eigenspinor of $h_0$ corresponding
 to an eigenvalue different from $\varepsilon_1$ and $\varepsilon_2$ will
 give rise to an eigenspinor of $h_1$. The eigenspinors of $h_1$
 related to the eigenvalues $\varepsilon_1$ and $\varepsilon_2$ will be
 obtained through $v \equiv (u^{\dagger})^{-1}$, that is $h_1v=v\lambda.$

\section{Factorization properties of Dirac  Hamiltonians and second order 
 supersymmetry}
\label{section3}

Let us here consider in addition to $L$ given in Eq.~(\ref{eq24}), 
its adjoint $L^{\dagger}$ defined by
\be
L^{\dagger}=-\frac{d}{dx} - (u_x u^{-1})^{\dagger}.
\label{eq29}
\ee
%
It satisfies an intertwining relation similar to Eq.~(\ref{eq9})
\be
L^{\dagger}h_1 = h_0L^{\dagger}.
\label{eq30}
\ee
This relation means that the operator $L^{\dagger}$ realizes 
the transformation in the opposite direction, i.e., the application 
of $L^{\dagger}$ to the eigenspinors of $h_1$ gives us the eigenspinors 
of $h_0$. The operator $L^{\dagger}L$ is thus such that applied to the
 eigenspinors of $h_0$, it gives back these eigenspinors. By definition,
 this is nothing but the fact that $L^{\dagger}L$ is a symmetry operator
 of the initial Dirac equation $h_0 \psi = E \psi$. Since we limit
 ourselves to the one-dimensional stationary context, this implies
 that $L^{\dagger}L$ is a function of $h_0$. Moreover because
 $L^{\dagger}L$ is a second order differential (matricial) operator
 while $h_0$ is of the first order, $L^{\dagger}L$ is in fact a
 polynomial of second order in $h_0$. More precisely, after tedious
 calculations, one can be convinced that
\be
L^{\dagger}L = (h_0-\varepsilon_1)(h_0-\varepsilon_2)
\label{eq31}
\ee
while a similar result holds for $LL^{\dagger}$
\be
LL^{\dagger} = (h_1-\varepsilon_1)(h_1-\varepsilon_2).
\label{eq32}
\ee
If we now introduce the 4 by 4 matrices
\be
H \equiv \left( \begin{array}{ll}
h_0 & 0\\
0 & h_1\\
\end{array}\right), Q^{\dagger}=\left( \begin{array}{ll}
  0 & L^{\dagger}\\
0 & 0\\
\end{array}\right), Q=\left( \begin{array}{ll}
  0 & 0\\
L & 0\\
\end{array}\right)
\label{eq33}
\ee
the relations~(\ref{eq9}) and~(\ref{eq30})--(\ref{eq32}) 
can be reformulated as
\be
[Q,H]=[Q^{\dagger},H]=0 \; , \; \{Q,Q^{\dagger}\}\equiv 
QQ^{\dagger}+Q^{\dagger}Q=(H-\varepsilon_1)(H-\varepsilon_2)
\label{eq34}
\ee
while
\be
Q^2=(Q^{\dagger})^2=0.
\label{eq35}
\ee
Relations~(\ref{eq34})--(\ref{eq35}) are the ones of a quadratic deformation 
of the superalgebra $sqm(2)$ subtending the usual supersymmetric
 quantum mechanics \cite{3}. This quadratic superalgebra cannot be seen
 directly from the Dirac equation and therefore we associate it
 with a hidden supersymmetry. Let us also and finally notice that
 a superalgebra similar to the one of~(\ref{eq34})--(\ref{eq35}) can also 
be found
 in the non-relativistic context when considering second order
 Darboux transformations \cite{10}.

\section{Examples}
\label{section4}

Let us now turn to some examples and see how our method provides us
 new exactly solvable Dirac potentials from known ones.

\subsection{The free particle case}

We consider here the potential
\be
v_0(x)=m \sigma_1 \; , \; x \in  {\Bbb R}.
\label{eq36}
\ee
Note that it corresponds to an unusual -but convenient- realization
 (the usual one being associated with $v_0(x)=m\sigma_3$) of the
 Clifford algebra subtending the one-dimensional Dirac equation.
 As stated in~(\ref{eq20}), it is necessary to take account of two
 eigenspinors corresponding to~(\ref{eq36}). 
Let $u_1$ and $u_2$ defined by
\be
u_1=\left( \begin{array}{l}
  {\rm ch}(kx)+\frac{cE}{k}{\rm sh}(kx)\\
{\rm ch}(kx+2\alpha)+\frac{cE}{k}{\rm sh}(kx+2\alpha)\\
\end{array}\right),
u_2=\left( \begin{array}{l}
  -{\rm ch}(kx)\\
{\rm ch}(kx+2\alpha)\\
\end{array}\right), 
\label{eq37}
\ee
be such eigenfunctions (of respective eigenvalues $\varepsilon_1=E$ 
and $\varepsilon_2=-E$) with
\be
k=\sqrt{m^2-E^2}, \; e^{2\alpha}=\sqrt{\frac{m-k}{m+k}}, 
\; c={\rm constant}.
\label{eq38}
\ee
The unique constraint to take care of in order to apply our method is to have 
a nonvanishing determinant: ${\rm det} u \neq 0$. 
Here it is precisely given by
\be
{\rm det} u = \frac{1}{E}
\left[m+E {\rm ch}\left(2kx+2\alpha\right)+
\frac{E^2c}{k}{\rm sh}\left(2kx+2\alpha\right)\right] 
\equiv \frac{1}{E} \Delta
\label{eq39}
\ee
and the parameter $c$ is such that $|c| < k/E$ in order 
to satisfy this constraint. The result~(\ref{eq28}) 
then gives rise to the new exactly solvable potential $v_1$
\be
v_1(x)=\frac{2E^2c}{\Delta}\sigma_3 +\left(m-\frac{2k^2}{\Delta}
\right)\sigma_1
\label{eq40}
\ee
whose eigenspinors can be obtained from the application of $L$ defined
 in Eq.~(\ref{eq24}) 
to the eigenspinors of the free Dirac Hamiltonian. Notice that the
 potential $v_1(x)$ given in Eq.~(\ref{eq40}) reduces to the well known
 one-soliton scalar potential when $c=0$.

\subsection{The generalized Coulomb case}

Before going to this example, we would like to mention that the
 usual radial equation associated to the (3+1)-dimensional Dirac
 equation is included in our developments. Indeed, the standard
 radial equation when coupled to scalar $W(x)$ and vector
 $V(x)$ potentials is
\be
\left\{\frac{d}{dx}-\frac{k}{x}\sigma_3+\left[M+W(x)\right]
\sigma_1+i\left[E-V(x)\right]\sigma_2\right\}\psi(x)=0, \; x \in {\Bbb R}_0^+
\label{eq41}
\ee
where $M$ and $E$ are the mass and the energy of the particle while 
$k$ is related to the total angular momentum. Eq.~(\ref{eq41}) can also be
written as
\be
\left\{i\sigma_2 \frac{d}{dx}+
\frac{k}{x}\sigma_1+\left[M+W(x)\right]\sigma_3
-\left[E-V(x)\right]\right\}\psi(x)=0, 
\label{eq42}
\ee
which coincides with $h_0 \psi(x)=E \psi(x)$ with $h_0$ 
defined in Eq.~(\ref{eq7}) and
\be
v_{12}^0(x)=\frac{k}{x},\ v_{11}^0(x)=M+V(x)+W(x),\ v_{22}^0(x)=-M+V(x)-W(x).
\label{eq43}
\ee

Let us now turn to our example. It corresponds to the choices of 
Ref.~\cite{11}:
\be
V(x)= \frac{\alpha}{x}, \; W(x)=\frac{\beta}{x}.
\label{eq44}
\ee
We refer to this example as the generalized Coulomb one 
because the choice $(\alpha=\frac{1}{137}, \beta =0)$ 
leads to the standard Coulomb interaction.

Let $\psi (x)=\left( 
\psi_1(x),
\psi_2(x)\\
\right)^T$ 
be a solution of Eq.~(\ref{eq41}) or equivalently~(\ref{eq42}), 
when the interactions~(\ref{eq44}) are taken into account. 
Using standard developments, we easily find the solutions in terms of 
hypergeometric confluent functions
\begin{eqnarray}
\psi_1(x)&=&e^{-\lambda_n x} x^{\mu}
\Bigg[-n\ _1F_1(1-n,2\mu +1;2\lambda_n x)\nonumber\\
&&\hspace{1cm}\mbox{}-\left(-k+\frac{\alpha M}{\lambda_n}
+\frac{\beta E_n}{\lambda_n}\right)\ _1F_1(-n,2\mu +1;2\lambda_n x)\Bigg],
\label{eq45}\\
\psi_2(x)&=&-\frac{\lambda_n}{M+E_n}e^{-\lambda_n x} x^{\mu}
\Bigg[-n\ _1F_1(1-n,2\mu +1;2\lambda_n x)\nonumber\\
&&\hspace{1cm}\mbox{}+\left(-k+\frac{\alpha M}{\lambda_n}
+\frac{\beta E_n}{\lambda_n}\right)\ _1F_1(-n,2\mu +1;2\lambda_n x)\Bigg]
\label{eq46}, 
\end{eqnarray}
where the parameters $\lambda_n$ and $\mu$ are constrained by
\begin{eqnarray}
\lambda_n^2 &=& M^2-E_n^2,\label{eq47}\\
\mu^2&=&k^2+\beta^2-\alpha^2\label{eq48}
\end{eqnarray}
while the number $n$ is defined by
\be
n=-\left(\frac{\alpha E_n}{\lambda_n}+\frac{\beta M}{\lambda_n} + \mu\right).
\label{eq49}
\ee
%
This relation can be solved for  the energies $E_n$ as 
\be
E_n=\frac{-\alpha \beta
  \pm (n+\mu ) \sqrt{\alpha^2+(n+\mu )^2-\beta^2}}{[\alpha^2+(n+\mu )^2]}M,
\label{eq50}
\ee
the plus or minus sign, as well as the values taken by $n$, 
having possibly to be chosen in order to ensure  the 
square-integrability of $\psi_1(x)$ and $\psi_2(x)$.

The most straightforward way to apply our method is to choose
\be
u_1=\left.\left( \begin{array}{l}
  \psi_1(x)\\
\psi_2(x)\\
\end{array}\right)\right|_{n=0}, 
u_2=\left.\left( \begin{array}{l}
  \psi_1(x)\\
\psi_2(x)\\
\end{array}\right)\right|_{n=1}.
\ee

In order to avoid heavy notations, we rewrite these choices as
\be
u_1=\left( \begin{array}{l}
  x^{\mu} e^{-\lambda_0 x}\\
c_1 x^{\mu} e^{-\lambda_0 x}\\
\end{array}\right), u_2= \left( \begin{array}{l}
  x^{\mu} e^{-\lambda_1 x}(1-c_2 x)\\
c_1 x^{\mu} e^{-\lambda_1 x}(1-c_3 x)\\
\end{array}\right)
\label{eq52}
\ee
with $\lambda_0$ and $\lambda_1$ defined through Eqs.~(\ref{eq49})
and~(\ref{eq50}), while
\begin{eqnarray}
c_1&=&\frac{\mu -k}{\alpha - \beta},\\
c_2 &=& \frac{\lambda_1}{1+2\mu}
+\frac{(E_1+M)(\mu -k)}{(\alpha - \beta)(1+2\mu)},\\
c_3&=& \frac{\lambda_1}{1+2\mu}
+\frac{(M-E_1)(\alpha - \beta)}{(\mu -k)(1+2\mu)}.
\end{eqnarray}

Applying finally the result~(\ref{eq28}), we obtain a (new) 
exactly solvable potential of the type
\begin{eqnarray}
v_1(x)&=&\frac{\alpha}{x}
+\left[-M+\frac{(\varepsilon_1-\varepsilon_2)}{c_2-c_3}
\left(\frac{2}{x}-c_2-c_3\right)\right]\sigma_3\nonumber\\
&&\hspace{-0.5cm}\mbox{}
+\left\{-\frac{k}{x}+\frac{(\varepsilon_1-\varepsilon_2)}{c_2-c_3}
\left[\left(c_1-\frac{1}{c_1}\right)\frac{1}{x}
+\left(\frac{c_2}{c_1}-c_1c_3\right)\right]\right\}\sigma_1.
\label{eq56}
\end{eqnarray}
In other words, we obtain a shape-invariant potential with respect to $v_0(x)$.

A particular example corresponding to the choices
\be
\alpha = 1, \; \beta =-1, \; \mu = 1, \; k=1
\label{eq57}
\ee
can be useful to illustrate the results here. Indeed we have
\be
\lambda_0=0,\ \lambda_1=\frac{4}{5}M,\ c_1=0,\ c_2=\frac{4}{15}M,\ 
c_1c_3=\frac{8}{15}M. 
\label{eq58}
\ee
The corresponding energies are  
\be
\varepsilon_1\equiv E_0=M, \; \varepsilon_2\equiv E_1=-\frac{3}{5}M.
\label{eq59}
\ee
The resulting potential is given by Eq.~(\ref{eq56}), i.e.,
\be
v_1(x)=\frac{1}{x}
+\left(\frac{3M}{5}+\frac{1}{x}\right)\sigma_3
+\left(\frac{2}{x}-\frac{4M}{5}\right)\sigma_1.
\label{eq60}
\ee

One can then determine the operator $L$, as defined in Eq.~(\ref{eq24}), 
connecting the eigenfunctions related to 
$v_0(x)=(1/x)+(M-1/x)\sigma_3+(1/x)\sigma_1$ 
and to $v_1(x)$ given in Eq.~(\ref{eq60}), respectively. It is given by
\be
L=\left( \begin{array}{cc}
\frac{d}{dx}-\frac{1}{x} & \frac{2M}{5}-\frac{2}{x}\\
0 & \frac{d}{dx}+\frac{4M}{5}-\frac{2}{x}\\
\end{array}\right).
\ee
Due to its definition, it is clear that 
$Lu_1\equiv Lu_2=0$, with $u_1$ and $u_2$ of  
Eq.~(\ref{eq52}) with the values~(\ref{eq58}). The definition of $L$ also 
implies that, whenever applied to any of the functions $\psi(x)$, 
it will give the eigenfunctions 
corresponding to $v_1(x)$ as expressed in Eq.~(\ref{eq60}). 
For instance, for $n=2$, we have
\begin{eqnarray}
&&L \left[\left(-\frac{2}{25}\exp^{-\frac{3M}{5}x}\right)
\left( \begin{array}{c}
 50x-30Mx^2+3M^2x^3\\
3(-10Mx^2+3M^2x^3)\\
\end{array}\right)\right] \nonumber \\
&&=-\frac{6}{125}\exp^{-\frac{3M}{5}x}M^2x^2\left( \begin{array}{c}
 -10+3Mx\\
5+3Mx\\
\end{array}\right),
\end{eqnarray}
and one can directly check that this is a solution of the final 
equation $h_1 \psi(x)=E \psi(x)$ with $E=-\frac{4}{5}M$. 
The other values of $n$ ($=3,4,...$) evidently lead to similar results.

The last information we  mention here is the possibility 
of obtaining new exactly solvable potentials and not only 
shape-invariant ones. This situation arises for example when we choose
\be
u_1=\left.\left( \begin{array}{l}
  \psi_1(x)\\
\psi_2(x)\\
\end{array}\right)\right|_{n=1}, 
u_2=\left.\left( \begin{array}{l}
  \psi_1(x)\\
\psi_2(x)\\
\end{array}\right)\right|_{n=2}.
\ee
With the set of parameters~(\ref{eq57}), we obtain
\begin{eqnarray}
v_1(x)&=&\frac{1}{50x-15Mx^2+12M^2x^3}\nonumber \\
&&\hspace{-2cm}\left( \begin{array}{ll}
 100+90Mx-60M^2x^2 & 100-115Mx-27M^2x^2+12M^3x^3\\
100-115Mx-27M^2x^2+12M^3x^3 & -120Mx+84M^2x^2
\end{array}\right)\nonumber\\
\label{eq64}
\end{eqnarray}
which has a different shape with respect to $v_0(x)$ and to the $v_1(x)$ 
given in Eq.~(\ref{eq60}). 
Evidently, one can determine the eigenfunctions related 
to this potential $v_1(x)$ defined in Eq.~(\ref{eq64}) 
through the application of 
the corresponding Darboux operator $L$ on the eigenfunctions $\psi(x)$ of 
$h_0$. One can also proceed in a similar way with different values of $n$ 
and obtain families of new exactly solvable potentials $v_1(x)$ whose 
eigenfunctions will be known through the application of the ad-hoc Darboux 
operator on the solutions of the generalized Coulomb problem.\\

{\bf ACKNOWLEDGMENTS}

The work of B.F. S. was partially supported by a grant of the 
Russian Foundation for Basic 
Research and a ``bourse de s\' ejour scientifique" from the F.N.R.S., Belgium.
The work of N. D. and B. V.d.B. was supported by the 
Institut Interuniversitaire des Sciences 
Nucl\'eaires de Belgique.

\end{document}